\documentclass[10pt,oneside]{article}
\usepackage{geometry}
\usepackage{graphicx}
\usepackage{graphics}
\usepackage{amsfonts}
\usepackage{amsmath}
\usepackage[english]{babel} 
\usepackage{geometry} 
\usepackage{hyperref}
\usepackage{epstopdf}

\title{
	The Italian primary school-size distribution\\
 	and the city-size: a complex nexus
}

\author{Alessandro Belmonte$^{1}$, 
Riccardo Di Clemente$^{1, 2}$
\& Sergey V. Buldyrev$^{3}$\thanks{\mbox{Corresponding author: \emph{E-mail~address}:~buldyrev@yu.edu}}\\
\footnotesize
$^{1}$\textit{IMT Institute for Advanced Studies Lucca, Piazza S. Ponziano 6, 55100, Lucca, Italy}\\
\footnotesize
$^{2}$\textit{Istituto dei Sistemi Complessi - CNR, Via dei Taurini 19, 00185 Rome, Italy}\\
\footnotesize
$^{3}$\textit{Department of Physics, Yeshiva University, New York, NY 10033 USA}
}

\begin{document}
\date{Published 23-June-2014 on Scientific Reports 4, 5301  \href{http://dx.doi.org/10.1038/srep05301}{DOI: 10.1038/srep05301} (2014)}

\maketitle

\begin{abstract}
We characterize the statistical law according to which Italian primary
school-size distributes.  We find that the school-size can be
approximated by a log-normal distribution, with a fat lower tail that
collects a large number of very small schools.  The upper tail of the
school-size distribution decreases exponentially and the growth rates
are distributed with a Laplace PDF. These distributions are similar to
those observed for firms and are consistent with a Bose-Einstein
preferential attachment process.  The body of the distribution features
a bimodal shape suggesting some source of heterogeneity in the school
organization that we uncover by an in-depth analysis of the relation
between schools-size and city-size.  We propose a novel cluster
methodology and a new spatial interaction approach among schools which
outline the variety of policies implemented in Italy.  Different
regional policies are also discussed shedding lights on the relation
between policy and geographical features.
\end{abstract}

\section*{Introduction}

There is a growing literature that nowadays sheds light on complexity
features of social systems.  Notable examples are firms and cities
\cite{gabaix09,gabaix99,allen1997cities,amaral98}, but many others
have been proposed \cite{byrne2002complexity,caves98}.  These systems
are perpetually out of balance, where anything can happen within
well-defined statistical laws \cite{bak1997nature, kauffman1996home}.
Italian schools system seems to not escape from the same
characterization and destiny.  Despite several attempts of the Italian
Ministry of education to reduce the class-size to comply with
requirements stated by law \cite{decret1, decret2, Belmonte2013}, no
improvements have been made and still heterogeneity naturally keeps
featuring the size distribution of the Italian primary schools.

In this paper we characterize the statistical law according to which
the size of the Italian primary schools distributes.  Using a database
provided by the Italian Ministry of education in 2010 we show that the
Italian primary school-size approximately distributes (in terms of
students) as a log-normal distribution, with a fat lower tail that
collects a large number of very small schools.  Similarly to the
firm-size \cite{dewit05,fu2005growth}, we also find the upper tail to
decrease exponentially.  Moreover, the distribution of the school
growth rates are distributed with a Laplassian PDF. These distributions
are consistent with the Bose-Einstein preferential attachment
process. These results are found both at a provincial level and
aggregate up to a national level, i.e. they are universal and do not
depend on the geographic area.

The body of the distribution features a bimodal shape suggesting some
source of heterogeneity in the school organization. We conclude that
the bimodality of the Italian primary school-size distribution is very
likely to be due to a mixture of two laws governing small schools in
the countryside and bigger ones in the cities, respectively.  The
bimodality source is studied in the paper by investigating the complex
link between schools and comuni, the smallest administrative centers
in Italy, addressed by the introduction of a new binning methodology
and a new spatial interaction analysis.

Several examples of different regional schooling organizations are
analyzed and discussed.  We use GPS code positions for schools in two
very different Italian Regions: Abruzzo and Tuscany.  We introduce a
measure of the average spatial interaction intensity between a school
and the surrounding ones.  We show that in regions like Abruzzo, that
are mainly countryside, a policy favoring small schools uniformly
distributed across small comuni has been implemented.  Abruzzo small
schools are generally located in low density populated zones, in
correspondence of very small comuni.  They are also very likely to
have another small school as closest and the median distance between
them is 8 $km$ that is also the distance between small comuni.  In
Tuscany, a flatter region with a very densely populated zone
along the metropolitan area composed by Florence, Pisa and Livorno, we
conversely find $1)$ a higher school density; $2)$ a stronger
interaction between small and big schools; $3)$ a greater average
proximity among schools.  We address these stylized facts by arguing
that the Italian primary school organization is basically the result
of a random process in the school choice made by the parents.  Primary
education is not felt so much determinant to drive housing choice,
like in US, because of the absence of any territorial constraint in
school choice.  Even if there is a certain mobility \emph{within} a
comune toward the most appealing schools, primary students generally
do not move \emph{across} comuni to attend a school.  As a result,
school density and school-size are prevalently driven by the
population density and then by the geographical features of the
territory.  This generates a mixture in the schooling organization
that turns into a bimodal shape distribution.


\section*{Results}

\subsection*{Empirical evidence}

We analyze a database on the primary school-size distribution in Italy that provides information on public and private schools, locations, and the number of classes and students enrolled.
Data are collected, at the beginning of every academic year, by the Italian Ministry of education to be used for official notices.
Our dataset covers $N = 17187$ primary schools in 2010 of which $91.31\%$ were public.
Almost four thousands are located in mountain territories, (which represent more than $20\%$) and $4101$ are spread among administrative centers (provincial head-towns). 

In Italy primary education is compulsory for children aged from six to
ten. However, the parents are allowed to choose any school which they
prefer, not necessarily the school closest to their
home, \cite{decret3}.  We define $x_i$ the size of the school $i \in
[1, \dots, N]$ as the number of students enrolled in each school.
Fig. 1(a) shows the histogram of the logarithm of the size of all
primary schools in Italy.  The red solid curve is the log-normal fit
to the data
\begin{equation}
P(\ln x)=\exp\left(-\frac{(\ln x-\hat{\mu})^2}{2\hat{\sigma}^2}\right) \frac{1}{\sqrt{2\pi}\hat{\sigma}}
\end{equation}
using the estimated parameters $\hat{\mu} = 4.77$ ($\hat{\mu}/\ln(10)=2.07$), the mean of the
$\ln x$ of the number of students per school, and its standard
deviation, $\hat{\sigma} = 0.85$ ($\hat{\sigma}/\ln(10)=0.37$).  On a non-logarithmic scale,
$\exp(\hat{\mu}) = 118$ and $\exp(\hat{\sigma}) = 2.34$ are called the
location parameter and the scale parameter, respectively
\cite{pitman1939estimation}.  The histogram in Fig. 1(a) suggests that
log-normal fits data quite well.  However, even a quick glance reveals
that there are too many schools with a small dimension and much less
mass in the upper tail with respect to the fit, suggesting that the
number of students of the largest schools is smaller than would be the
case for a true log-normal.  In other words, similarly with firms-size
distribution \cite{stanley95}, tails seem to distribute differently
from the log-normal distribution.  Also Fig. 1(a) reveals a bimodal
shape of the school-size distribution that we will extensively
investigate below.

\begin{figure}[!hbt]
\includegraphics[width=.99\textwidth]{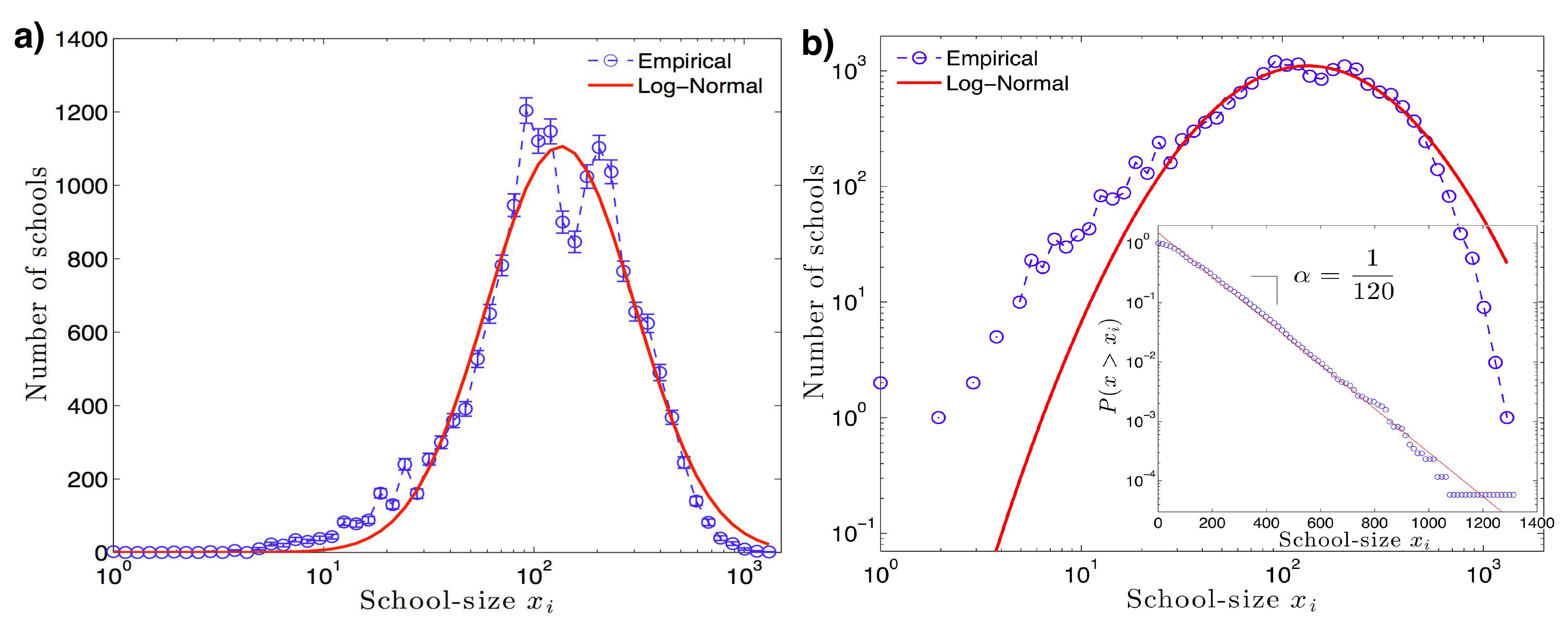}
\caption{ \textbf{School-size distribution. a.}
	Italian primary school-size distribution according to the number $x_i$ of student per school $i \in [1, \dots, N]$ for the year 2010.
	The empirical distribution is drawn in blue (each circle is a bin);
	the red line stands for the Gaussian fit with mean $\hat{\mu} = 4.77$ ($\hat{\mu}/\ln(10) = 2.07$) and standard deviation $\hat{\sigma} = 0.85$ ($\hat{\sigma}/\ln(10)=0.37$).  On a non-logarithmic scale, $\exp(\hat{\mu}) = 118$ and $\exp(\hat{\sigma}) = 2.34$.
	$N = 17187$.
	Statistical errors (SE) are drawn in correspondence of each bin, according to $\sqrt{N_{bin}}$.
	SE  are bigger in the body of the distribution and tinier in the tails.
	Nevertheless, central bins space from the two peaks, $m_1 = 1.7$ and $m_2 = 2.3$, at least $6$ times the SE, equals on average to $\sqrt{10^3} = 32$. 
	In this case the probability to have a non bimodal shape under our distribution is pretty narrowed and tends to $[\frac{1}{6}e^{-\sigma^{2}/2}]^{2}\approx 10^{-17}$. 
			\textbf{b.}
	Italian primary school-size distribution in log-log scale.
	As expected, the theoretical distribution has drawn as a perfect parabola (the red curve), $y = ax^2 + bx + c$, such that $\hat{\mu}= -b/2a$ and $\hat{\sigma} = - 1/2a$. 
Conversely, the empirical distribution does not plot as a parabola, at least for what regards to the tails which deviate from the log-normal.
	The inset figure shows a functional form of the right tail of the empirical distribution.
	We plot the cumulative distribution, $P(X > x_{i}) = \exp(-\alpha x_i)$, of school sizes in semi-logarithmic scale with characteristics size $\alpha = 0.0084$. 
	This in turn means that there are approximately 120 students per school.}

\end{figure}

These findings can be detected in a more powerful way by plotting the
histogram in a double logarithmic scale, comparing the tails of the
log-normal distribution with those of the empirical one.  We do this
in Fig. 1(b) where y-axes represents the logarithm of the number of
schools in the bins whereas in the x-axes the logarithm of the number
of students stands.  The empirical distribution differs significantly from the theoretical 
distribution which is a perfect parabola (the red curve), both in the tails 
and in the central bimodal part.
A functional form of the right tail of the empirical distribution is
revealed in the inset of Fig. 1(b) where we plot the cumulative distribution
$P(X > x)$ of school sizes in semi-logarithmic scale. The straight line fit
suggests that the right tail decreases exponentially $P(X > x)= \exp(-x\alpha)$ 
with a characteristics size $\alpha = \frac{1}{120}$. This in turn means
that there are approximately $120$ students per school and also that the
distribution of large schools declines exponentially. The exponential decay of the right tail of 
size distribution is consistent with Bose-Einstein preferential attachment process and is observed
in the distribution of sizes of universities and firms.

\begin{figure}[!hbt]
\includegraphics[width=.99\textwidth]{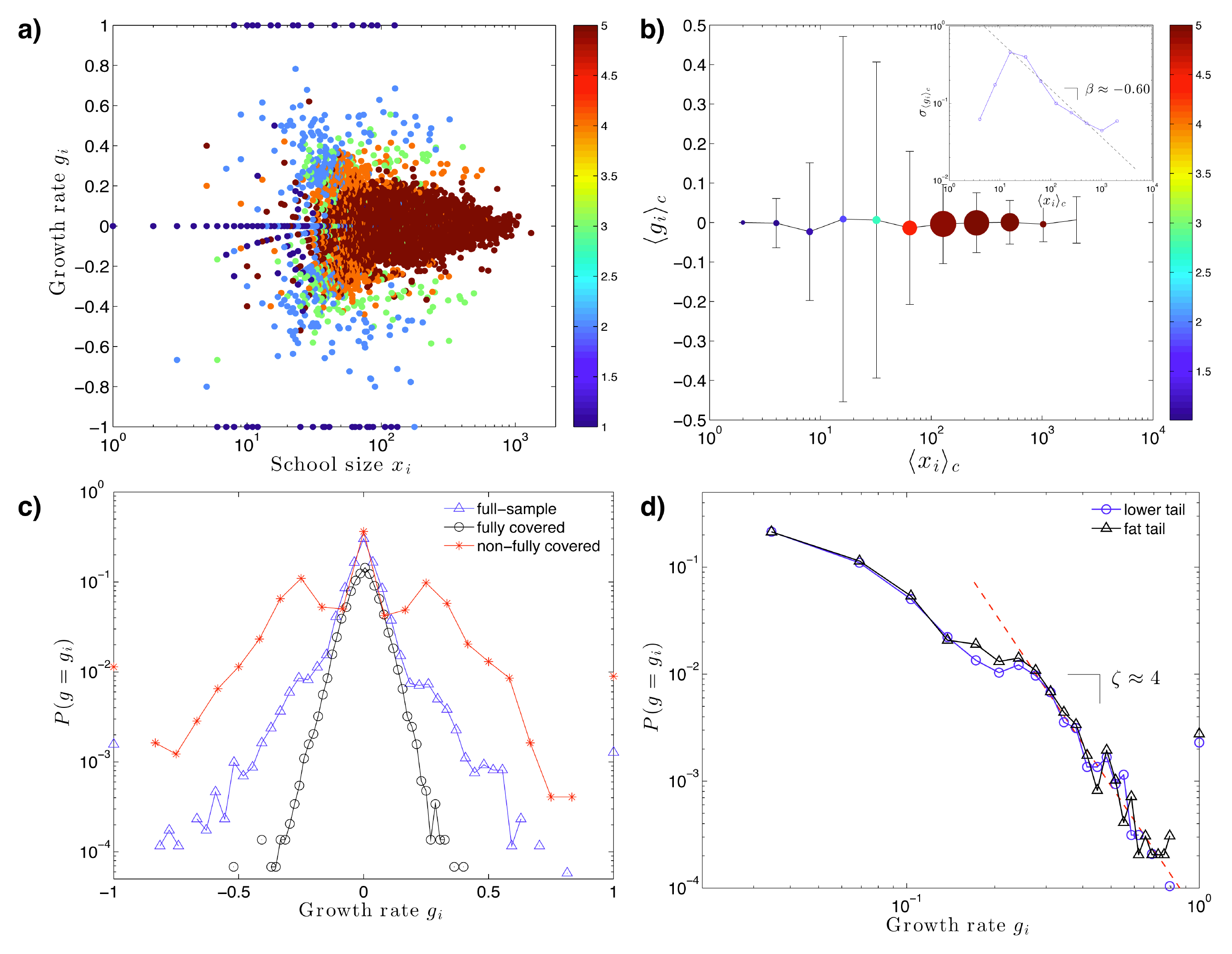}
\caption{\textbf{The growth rate distribution of the Italian primary schools in 2010.}
	The growth rate $g_i$ is defined according to Eq. (\ref{g}).
			\textbf{a.} The growth rate and school-size relationship. 
Colors, according to the vertical bar on the right-hand side of the graph, are the number of grades $J_i$ provided by the school $i$. 
Smaller schools (in blue) with $J_{i}=1$ are both the newest one (just created, with $\lambda = 1$) and schools that are going to close (with $\mu = 1$).
They can also be schools that do not grow yet providing just one grade (i.e. $j = 3$). 
	\textbf{b.} The mean growth rate clusters around zero across different subsets $c$ that are differently populated by $n_c$ schools according to the size of the circles.
The color of the circles stand for the average number of grades $J_i$ (the same gradient color bar of Fig. 2(a) is used here).
The variability within each cluster $c$ is shown in the inset figure.
Apart from schools with $x_i < 10$, namely hospital-based schools mostly similar to one another, the standard deviation is found to be decreasing with school-size by a rate of $\beta \approx .60$.		
	\textbf{c.} The probability density function $P(g = g_i)$ of growth rate has been plotted underlying a Laplace PDF in the body around $P(g) = 1$ and $P(g) \approx 10^{-1.5}$.
Blue triangles ($\triangle$) stand for the full sample distribution, black circles ($\circ$) indicate mature schools with $J_i = 5$, and red stars ($\ast$) schools with $J_i = 1$.			
	\textbf{d.} The plot reports empirical tests for the tails parts of the PDF of growth rate, the upper one in blue ($\circ$), and the lower one in black ($\square$).
The asymptotic behavior of $g$ can be well approximated by power laws with exponents $\zeta \approx 4$ (the magenta dashed line).}
\end{figure}

Next we investigate the growth rates of elementary schools.  Since
temporal data are not currently available, we look at the single
academic year, the 2010, and define the growth rate $g_i$ as follows:
\begin{equation}
\label{g}
	g_i \equiv \frac{x_{i}^{1} - x_{i}^{5}}{\sum_{j = 1}^{5}x_{i}^{j}} = \lambda_i - \mu_i,
\end{equation}
where $x^{j}_{i}$ stands for the number of students attending the
$j\mbox{\emph{-th}}$ grade in school $i$, with $j \in [1,5]$;
$\lambda_{i} \equiv x_{i}^{1}/\sum_{j = 1}^{5}x_{i}^{j}$ is the
fraction of students that have been enrolled in the first grade at six
years old in school $i$, whereas $\mu_{i} \equiv x_{i}^{5}/\sum_{j =
  1}^{5}x_{i}^{j}$ is the fraction of students that exit the school
after the $5\mbox{\emph{-th}}$ grade.  Fig. 2(a) shows the relation
between growth rate $g_i$ and school-size $x_i$.  The numbers of
grades $j$ provided by each school $i$, named $J_{i}$, is defined by
the color gradient bar on the right side of the Fig. 2(a).  Blue
circles identify schools with $J_i = 1$.  Such a group collects
schools just established only providing the $1\mbox{\emph{-st}}$
grade, i.e. with $\lambda_i = 1$ \emph{and} $\mu_i = 0$, or that are
going to close providing only the $5\mbox{\emph{-th}}$ grade,
i.e. with $\mu_i = 1$ \emph{and} $\lambda_i = 0$.  As soon as more
grades are provided (colors switching to the warm side of the bar)
schools tend to cluster around a null growth rate.

In Fig. 2(b) we investigate the growth/size relationship in depth.  We
demonstrate the applicability of the Gibrat law that states that the
average growth rate is independent on the size \cite{gibrat,
  sutton97}.  We define the average of the school size in each bin $c$
as $\langle x_{i} \rangle_{c}$.  The number of school in each bin
$n_{c}$ is represented by the size of the circle and the average
number of grades $\langle J_i\rangle_{c}$ is depicted according to the
color gradient on the right side (the same of Fig. 2(a)).
Independently from the size and the number of grades provided, schools
do not grow on average.  Nevertheless, we find more variability in
smaller schools, apart from schools with $x_i < 10$, namely
hospital-based schools mostly similar to one another, and the standard
deviation of the growth rate $\sigma_{g(\langle x_{i} \rangle_{c})}$
is found to be decreasing as $\langle x_{i} \rangle_{c}^{-\beta}$ with
school-size by a rate of $\beta \approx .60$ (subFig. 2(b) inset).
This is consistent with what has been found for other complex systems
like firms or cities \cite{fu2005growth, fu2007generalized,
  pammolli2007generalized, axtell01, growiec07, stanley96}.

In Fig. 2(c) we study the growth rate distribution, where the
probability density function $P(g = g_i)$ of growth rate has been
plotted.  The blue line represents the full sample (all the schools)
distribution.  Black and red colors identify the full capacity schools
($J_i = 5$) and the schools with $J_i < 5$, respectively.  Regardless
of the number of grades provided, the growth distribution underlines a
Laplace PDF in the central part of the sample
\cite{ayebo2003asymmetric}.  The not-fully covered schools show a
three peak behavior, where the left peak represents schools which are
going to close, the central peak gathers schools that provide several
grades but still in equilibrium phase, and the right peak is made up
by the growing schools.  Fig. 2(d) reports empirical tests for the
tails of the PDF of the growth rate of the full sample (the upper one
in blue, and the lower one in black).  The asymptotic behavior of $g$
can be well approximated by power laws with exponents $\zeta \approx
4$ (the magenta dashed line), bringing support to the hypothesis of a
stable dynamics of the process \cite{pammolli2007generalized}.  All
these findings are consistent with the Bose-Einstein process according
to which the size distribution has an exponential right tail, a
tent-shaped distributed growth rate $g_i$, with a Laplace cap and
power law tails, the average growth rate is independent of the size, and the
size-variance relationship is governed by the power law behavior with
exponent $\beta \approx 0.5$ \cite{bookbul}.

\subsection*{City size and school size}

Fig. 1(a) features the coexistence of two peaks, the first peak
corresponding to $\log_{10} x_{i} \equiv m_{1} = 1.7$ and the second
one to $\log_{10} x_{i} \equiv m_2 = 2.3$, divided by a splitting
point in correspondence of $\log_{10} x_{i} \equiv \bar{m} \approx
2.1$. The school sizes corresponding to these features are $\mu_1=10^{m_1}=50$,
$\mu_2=10^{m_2}=200$, and $\bar{\mu}=10^{\bar{m}}=128$, with $\bar{\mu}$ approximately
equal to the average school size.  $39\%$ of the Italian primary schools distribute on the right
of $\bar{\mu}$, and more than 60\% distribute on the left side.  We
test the alternative hypothesis of unimodality by looking at the
probability that the numbers of schools in the two central bins
$n_1,n_2$ are not smaller and the numbers of schools in the next three
bins $n_3,n_4,n_5$ are not larger than a certain number $
n^\ast$ provided that the standard deviation of the number of
schools in these bins due to small statistics is $\sqrt{
  n^\ast}$.  This probability is equal to $p(n^\ast)=\prod_i {\rm
  erfc}(|n_i-n^\ast|/\sqrt{2n^\ast})/2$ and it reaches maximum $p_{\max}\approx
4 \times 10^{-15}$ at $n^\ast=980$. Accordingly, we establish the bimodality
with a very high confidence. This is also consistent with the
bimodality index that we find to be equal to $\delta = (\mu_1 -
\mu_2)/\sigma = .45$, \cite{wang09}.

In this section we investigate the source of this heterogeneity that
we find to be related to geographical and political features of the
country and remarkably on the size of the comuni, the smallest
administrative centers in Italy (information on comuni are provided
by the Italian statistical institute, ISTAT), also here referred
interchangeably as cities regardless of the size, $p_k$.

\begin{figure}[!hbt]
\includegraphics[width=.99\textwidth]{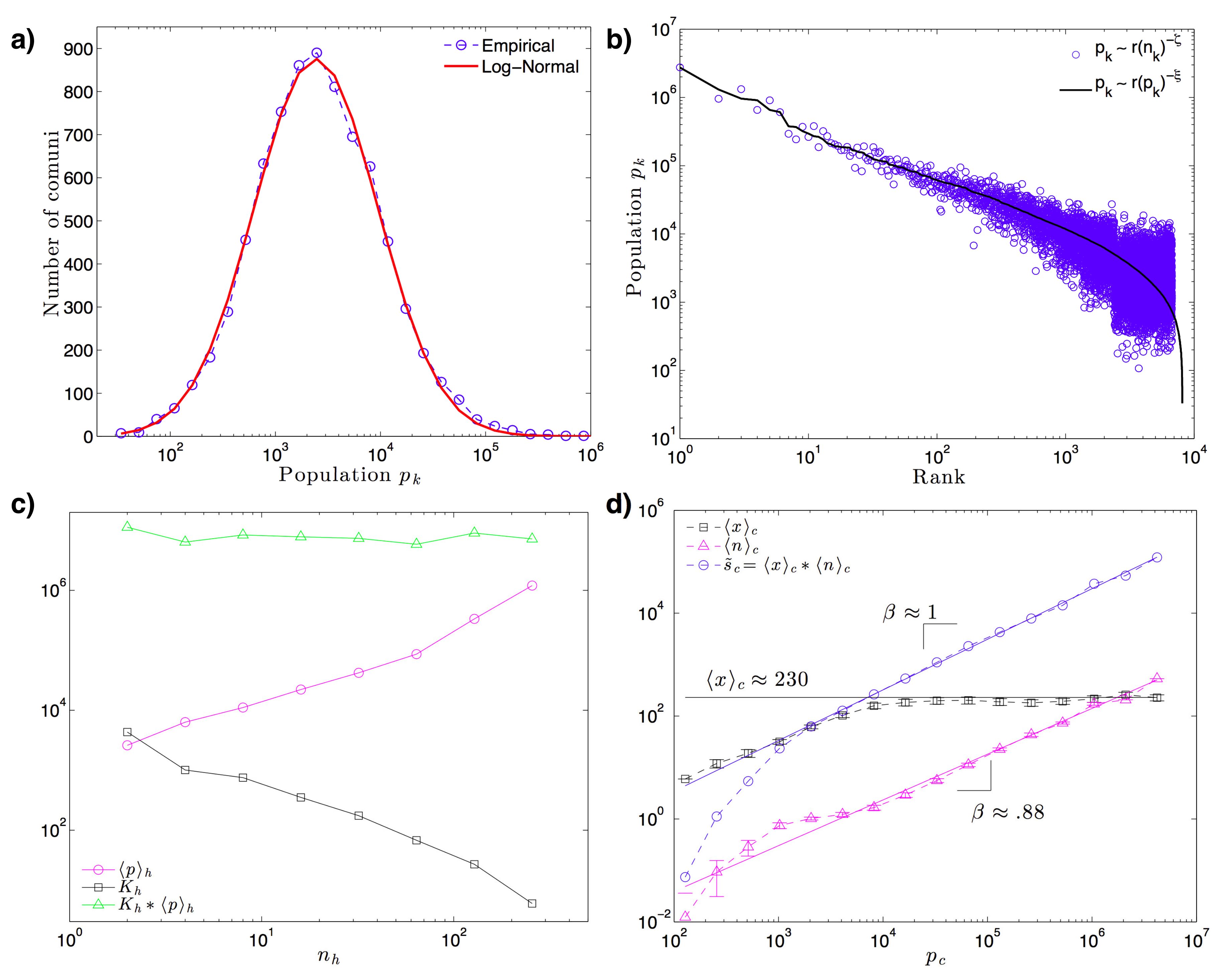}
\caption{ \textbf{Population and cities features.\label{popunif}}			
	\textbf{a.} The Italian city-size distribution for $K = 8092$ observations.
Blue circles stand for each city-bin whereas the red solid line draws the log-normal fit of the data.
Conversely to the school-size distribution depicted in Fig. 1(a), the city-size PDF features single-peakedness, but similarly it has a power-law decay in the upper tail. 			
	\textbf{b.} Zipf plot for Italian cities according to the size $p_k$ and the number of schools  $n_k$.
The black line draws the classical Zipf plot $p_k \sim r(p_k)^{-\xi}$, with cities ranked according to population $p_k$.
Blue circles instead depict the Zipf plot $p_k \sim r(n_k)^{-\zeta}$, with cities ranked according to the number of schools $n_k$.
Consequently, the sample reduces to $M = 6726$ over $N = 8092$ since more of the $15\%$ of the cities have no schools.		
	\textbf{c.} Each comune is assigned to 8 clusters, according to Eq. \ref{clusterschool}, and scattered against population, the magenta line ($\circ$) and the number of cities $K_h$, the black line ($\square$).
The interaction term, $K_h * \langle p \rangle_h$, the green line ($\triangle$), represents the total population living in each city-cluster $h$.
	\textbf{d.} According to Eq. \ref{clustercomuni} $K$ cities are assigned to $C = 16$ clusters.
In the x-axis the number of inhabitants in cluster $c = \{7, 22\}$ is scattered against the average number of schools (magenta line ($\triangle$)) and the average school-size $\langle x\rangle_c$ (the black line ($\square$)).
The interaction term ($\circ$), representing the typical number of schooling-aged population in cluster $c$, $\tilde {s_{c}} = \langle x\rangle_{c}*\langle n\rangle_{c}$ distributes as a power law with coefficient $\beta \approx 1$ for cities bigger than $10^3$ inhabitants, and it is drawn in green.
For smaller comuni, instead, the line drops meaning that a smaller fraction of young people features them.}
\end{figure}

In 2010, $K = 8092$ comuni have been counted in Italy, the $40\%$ of
which located in the mountains $\mathcal{M}$.  Each city $k \in [1,
  \dots, K]$ has $n_k \geq 0$ schools (more than $15\%$ of the cities
have no schools) and population $p_k$, which distributes approximately
as a log-normal PDF (see Fig. 3(a)), except for the right tail that is
distributed according to a Zipf law, i.e. $p_k \sim r(p_k)^{-\xi}$
with slope $\xi\approx 1$ \cite{gabaix99,allen1997cities,lauset2009power,
  eeck04,reed01}.  In Fig. 3(b) we find $\xi \approx .80$, in Italy,
that is exactly the slope of the power law $p_k \sim r(n_k)^{-\zeta}$
which links the population $p_k$ with the rank of this city in terms of
 number of schools $n_k$ (blue circles in Fig. 3b),
i.e. $\zeta = \xi \approx .80$. This means that the first city, Rome, has almost the double number
of schools than Milan, and triple of Naples, while Rome has almost
the double of inhabitants of Milan, and the triple of Naples.
This amounts to say that $n_k$ is a good proxy for the city-size.

We use the number of schools to assign comuni to different clusters $h \in [1, \dots, H]$, according to 
\begin{equation}
	h = \{\forall \, k \in [1, \dots, K] : \, 2^{h-1} \leq n_{k} < 2^{h}\}.
	\label{clusterschool}
\end{equation}
Accordingly, the first bin $h = 1$ gathers all the comuni with only
one school; the second one collects all the comuni with $n_k = [2,
  3]$, and so on.  Though we find the average population $\langle
p\rangle_h$ to increase across different city-clusters $h$, less
comuni $K_h$ lie in more populated clusters (the magenta and black
lines in Fig. 3(c)).  Interestingly, we find the interaction term $K_h
\langle p \rangle_h$, the green line in Fig. 3(c), to distribute
uniformly across different comuni-clusters, meaning that in small
comuni with $n_k = 1$ live the same population than in bigger ones
with much more schools.

Nevertheless, population is differently composed across city-clusters
and a smaller fraction of young people is found in smaller comuni.
To see that we also introduce a clusterization of comuni according to
population.  Each comune is assigned to a cluster $c \in [1, \dots,
  C]$ composed by all the comuni $k$ with population $p_k$ ranging
from $\psi^{c-1}$ to $\psi^{c}$, i.e.
\begin{equation}
	c = \{\forall \, k \in [1, \dots, K] : \, \psi^{c-1} < p_{k} \leq \psi^{c}\}.
	\label{clustercomuni}
\end{equation} 
Setting the parameter $\psi = 2$ yields $C = 23$ clusters.  Although
the first seven sets are empty because no comuni in Italy has less
than $128$ inhabitants, the first (non-empty) cluster, $c = 8$, collects
very small comuni with $p_k \in (128, 256]$. The last one, $c = 23$,
  conversely, is composed by the biggest cities with $p_k \in (2^{22},
  2^{23}]$.  In Fig. 3(d) we plot the average number of schools $\langle n\rangle_{c}$
(magenta line) and the average school-size $\langle x\rangle_{c}$ (the
blue line) against the comuni size $p_c$ for each non-empty cluster $c$.
We find that the average number of schools increases
as a power law with coefficient $\beta=0.88$. This is consistent with the literature
\cite{gabaix99,allen1997cities,lauset2009power, eeck04,reed01} that
has stressed the emergence of scale-invariant laws that characterize
the city-size distribution. The average school-size
increases with the population of the city reaching an asymptotic value at $\langle
x\rangle_{c} \simeq 230$ students per school in the large cities.  As expected, the
interaction term, representing the average number of school-aged
population in comuni belonging to cluster $c$, $\tilde {s}_{c} = \langle
x\rangle_{c}*\langle n\rangle_{c}$, behaves linearly with the comuni size
except for small comuni with $p_c < 10^3$, for which the school-aged population constitutes a smaller fraction
of the total population than in large cities.  

\bigskip
\begin{figure}[!hbt]
\includegraphics[width=.99\textwidth]{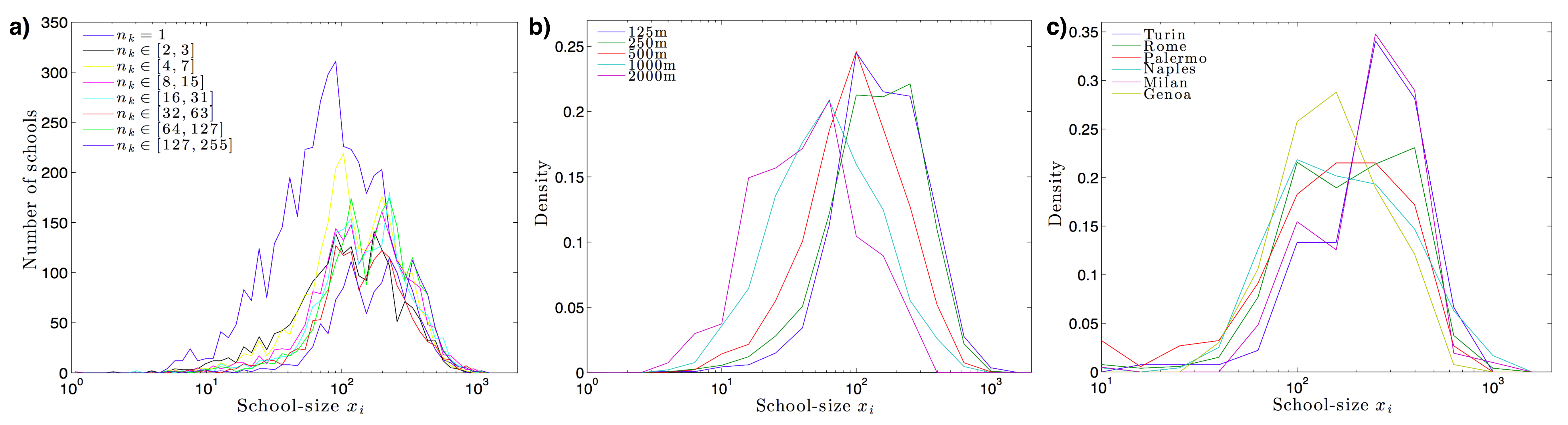}
\caption{\textbf{School-size distribution conditional on comuni features.}
			\textbf{a.} 
	School-size distribution for different city-samples clustered according to the number of schools, i.e. to Eq. \ref{clusterschool}.
	Only comuni with $n_k = 1$ show a single peak  school-size distribution, clustered around $m_1$ (the blue line on the top).
	They have an average population of $2000$ inhabitants and the $81\%$ are located in mountain territories.
			\textbf{b.} 
	School-size distribution for different city-samples clustered according to the altitude.
	The altitude of the comune shift the school-size distribution (shift location effect) as higher comuni are generally smaller schools. 
			\textbf{c.} 
	School-size distribution in the six biggest Italian cities.
	Except in Rome, the hypothesis of unimodality may not be reject none of the biggest cities, and find geography to drive the size of the schools.
	In particular, flatter cities, such as Milano and Torino, mostly contribute to second mode $m_2$, whereas in Genova, Italian city built upon mountains that steeply ended on the see, all the school-size distribution stands on the left side.}
	\end{figure}

In Fig. 4 we investigate the school-size distribution according to the comuni features.
To this end, Fig. 4(a) draws the distributions of $\log_{10} x_i$ conditionally on the number of schools, $n_k$, in the comune $k$.
It yields $8$ curves, one for each cluster $h$ defined in Eq. \ref{clusterschool}. 
The first cluster is drawn in blue distributing all the schools located in comuni where only one school is provided.
The black line distributes all the schools provided in comuni with two or three schools (i.e. $h = 2$);
and so on.
The interesting point of Fig. 4(a) is that only the school-size distribution of the smallest comuni (with $n_k = 1$) features a unimodal shape.
The reason for that relies on the fact that comuni with only one school are geographically similar:
they are the $57\%$ of the total, with little more than $2000$ inhabitants, the $81\%$ of which are located in mountain territories.

The relationship between school-size and altitude is investigated in Fig. 4(b), where comuni are assigned to different bins according to the altitude.
It yields 5 bins:
the first bin (drawn as a blue line) gathers all the comuni whose altitude is lower than $125$ meters above the see level (labeled $125$ in Fig. 4(b)).
Comuni with an altitude between $125$ and $250$ meters above the see level composed the second bin (the green line).
These two distributions cluster around the second mode $m_2$.
However, the greater the altitude of the comuni the more the school-size distributions of the different bins move left, mostly contributing to the first mode $m_1$.
Such a shift location effect is evident considering the comuni with an altitude between $250$ and $500$ meters above the see level (the red line), whose school-size distribute with roughly the same mean of the distribution in Fig. 1(a).
Higher comuni (the cyan and purple lines for comuni higher than $500$ and $1000$ meters respectively) clusterize around $m_1$.

Even the largest cities are very different from each other in terms of their school size distribution. This heterogeneity is very likely to be driven by geographical features.
We argue this point in Fig. 4(c), where we restrict our interest on the largest Italian cities belonging to cluster $h = 8$ (and to the first two bins in terms of altitude in Fig. 4(b)).
These cities provide a number of schools $n_k$ within $127$ and $255$, whose size distribution overall shows a three-peak shape (the bottom blue line in Fig. 4(a)).
By plotting the distribution by city we show that all the traces of bimodality disappear.
In particular flatter cities, such as Milano and Torino, mostly contribute to second mode $m_2$, whereas in Genova, an Italian city built upon mountains that steeply slope towards the sea, the school-size distribution is unimodal contributing mostly to the first mode $m_1$.

\bigskip

\begin{figure}[!hbt]
\includegraphics[width=.99\textwidth]{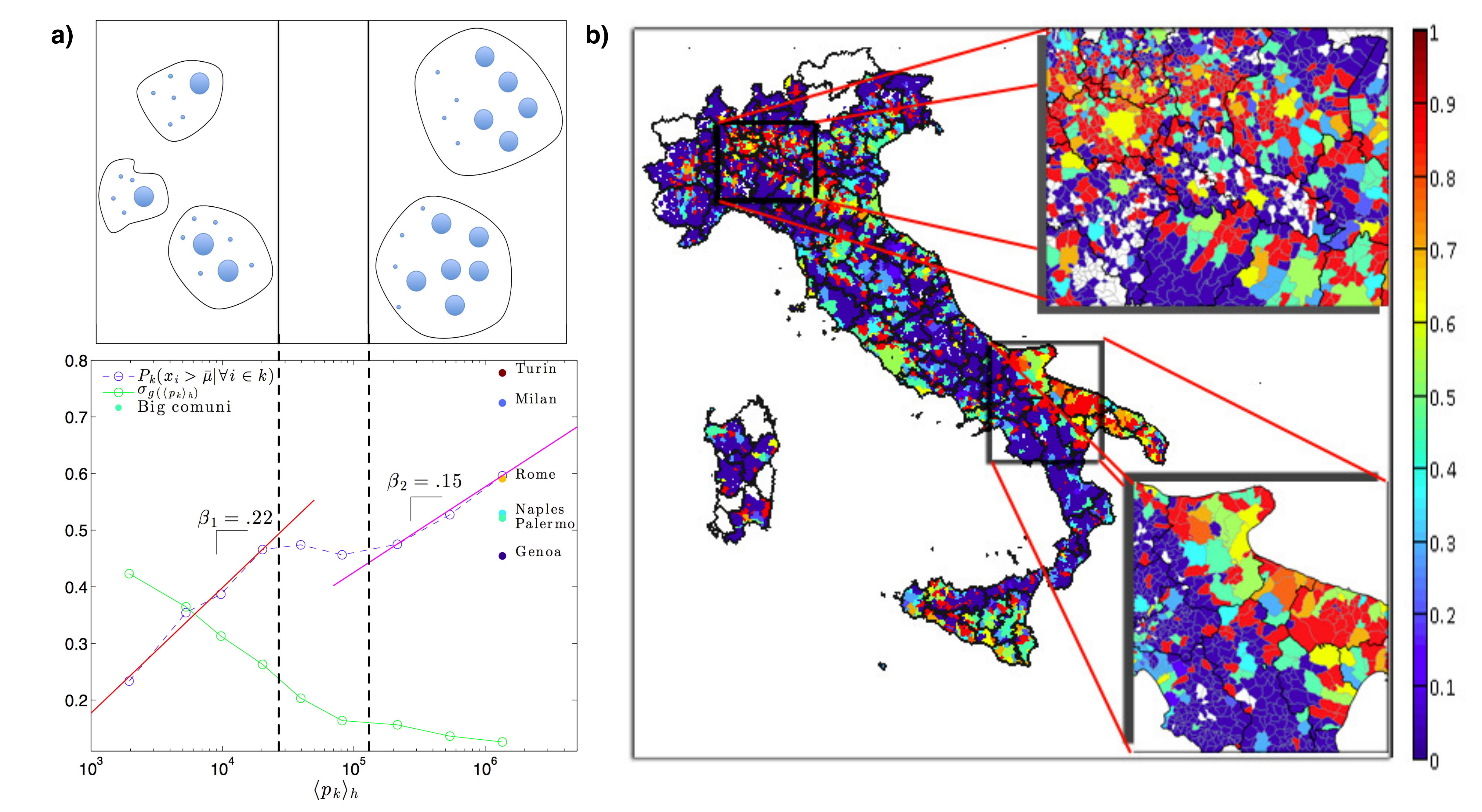}
\caption{\textbf{Fraction of large schools in comune $k$.}
	\textbf{a.}
	The panel above shows the process according to which each comune, with population $p_k$ defined by the size of the the black circles, is assigned to either patterns on the basis of the size of the schools provided in there (the small blue circles).
	The panel below shows that more populated clusters of cities are, on average, more likely to have schools sized around $m_2$.
	The relationship, depicted in blue, is however non monotonic.
	In correspondence of each bin $h$, the standard deviations has been computed, underlining the outstanding variability in very small cities (the green line).
	\textbf{b.}
	Spatial distribution of cities according to $P_k(x_i > \bar{\mu}| \forall \, i \in k)$.
	Warmer territories stand for cities more likely of having schools distributed around $m_2$.
	The two figure inset underline the region around Milan (in the North), on the top, and the regions of Basilicata (mostly mountain, at the left side) and of Apulia (mostly flat, at the right side), on the bottom. Maps generated with Matlab.}    
	\end{figure}

Another way to look at the effect of geography on the comunal school-size is to compute the fraction of large schools on the total within each comune $k$:
\begin{equation}
\label{frac}
	P_k(x_i > \bar{\mu}| \forall \, i \in k) \equiv \frac{n_k(x_i > \bar{\mu})}{n_k} \quad \forall \, i \in k,
\end{equation}
where $n_k(x_i > \bar{\mu})$ stands for the number of schools that, in each comune $k$, are larger than the minimum $\bar{\mu}$ of the school-size distribution shown in Fig. 1(a).
It can also be interpreted as the contribution rate of a comune $k$ to the second mode $m_2$.
The upper panel of Fig. 5(a) diagrammatically explains how $P_k(\cdot)$ is computed.

We firstly study the relationship between $P_k(\cdot)$ and population, then looking at the spatial distribution across the Italy.
In Fig. 5(a), we clusterize comuni according to Eq. \ref{clusterschool}, and for each bin $h$ we compute the average $\langle P_k(x_i > \bar{\mu}| \forall \, i \in k)\rangle_h$ and population $\langle p_k\rangle_h$.
Interestingly, the scatter shows how $P_k(\cdot)$ does not increase monotonically with population, showing the existence of two city-patterns.
More precisely, cities with less than $10^4$ inhabitants follow a pattern according to which the fraction of big schools, with $x_i > \bar{\mu}$, increases, on average, with population at a rate of $\beta_{1} \approx .22$;
in cities with more than $10^5$ we find the effect of population to be smaller, corresponding to $\beta_{2} \approx .15$.
Cities with population in between, i.e. $10^4 \leq p_k \leq 10^5$, lie in a critical state suggesting that exogenous shocks might lead a city to either patterns, make it more or less likely to contribute to the second mode $m_2$.

Overall, the distribution of $P_k(x_i > \bar{\mu}| \forall \, i \in k)$ is strongy correlated with the geographical features of the comuni territory.
The map in Fig. 5(b) clarifies this point; 
all the mountain territories, Apennines that represent the spine of the peninsula and the Alps on the northern side, turns to be comuni with small schools, since the share of small schools in mountain comuni is equal to $P(x_i \leq \bar{\mu}| k \in \mathcal{M}) = 0.72$.  
As soon as the probability to contribute to $m_2$ increases the colors get warmer; 
but this is very unlikely to be in mountain territories, because less than $30\%$ of mountain comuni
contribute to the antimode.  
Some regional patterns are also shown in the insets.
The first upper panel depicts the area around Milan, which is surrounded by warm colors that mostly dye the Pianura Padana around.
On the south side, Appennines approch and colors get blue with a lot of comuni with no schools (depicted in white).
This pattern is more evident in the lower panel, which maps the region of Apulia, flat and mostly red, and the Basilicata on the left side, mountainous and mostly blue colored.

\subsection*{Countryside versus dense regions}

\begin{figure}[!hbt]
\includegraphics[width=.99\textwidth]{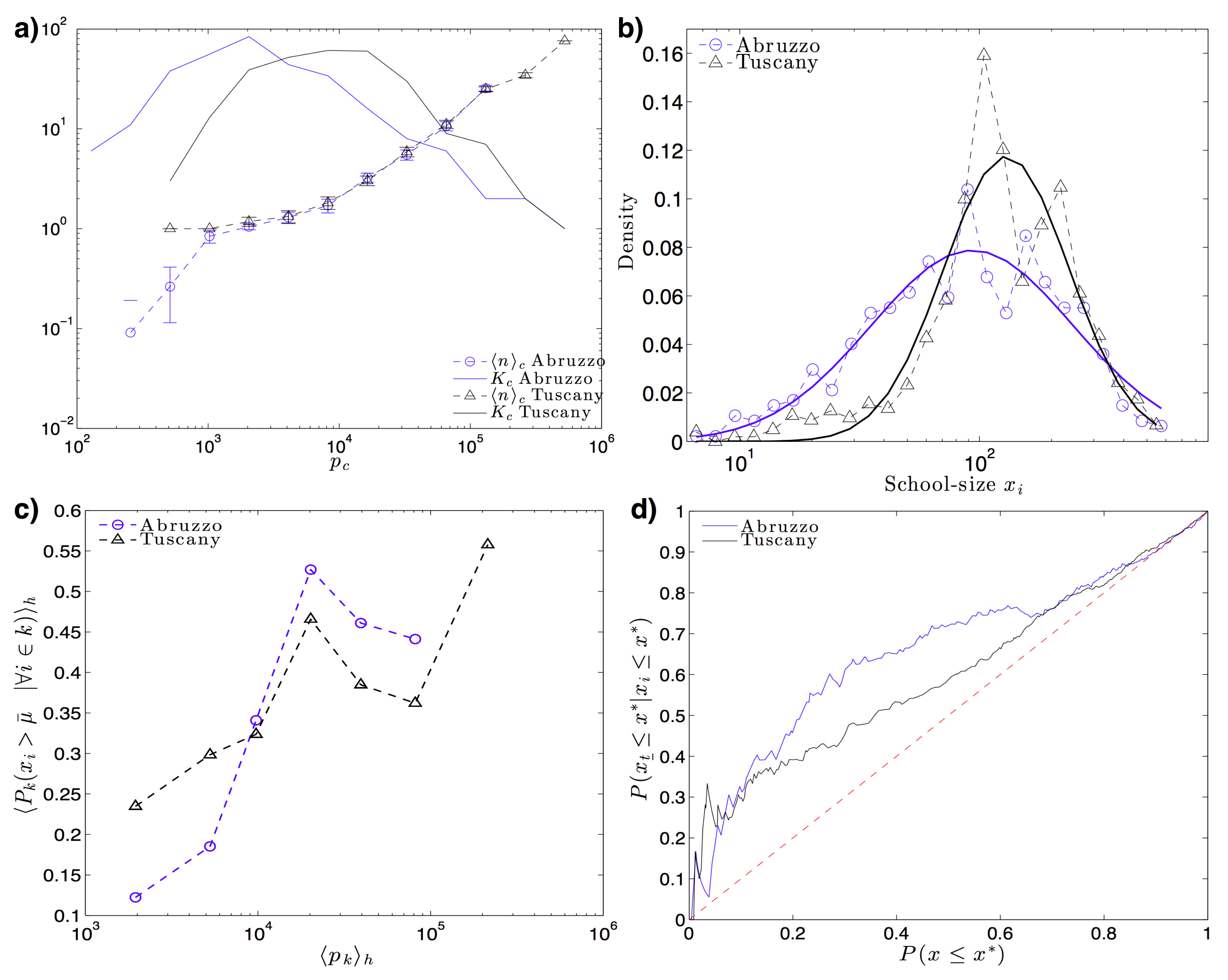}
\caption{ \textbf{Regional analysis.} 
	\textbf{a.} The figure distributes the city-size in Abruzzo (blue) and Tuscany (black) by plotting the number of comuni, $K_c$, against the number of inhabitants, $p_c$.
Also shown is the average number of schools in a comune in Abruzzo and Tuscany, belonging to a bin $c$ defined by Eq. \ref{clustercomuni}, by the circled- and triangled-connected lines respectively. 
	\textbf{b.}School-size distribution in Abruzzo (blue) and Tuscany (black).
Both pdf are approximately lognormal and bimodal with splitting point equal to $128$ and $151$ students per school respectively.		
	\textbf{c.}
 Average fraction of big schools in each comuni bin, defined by Eq. 3, in Abruzzo (blue $\circ$) and Tuscany (black $\triangle$).
	The plot shows that more populated comuni are, on average, more likely to have schools sized around $m_2$, in both regions.
Yet, in mountain regions, such as Abruzzo, smaller comuni have also smaller schools on average.
	\textbf{d.}
	The conditional probability is plotted in the y-axis, for an arbitrary school size $x^\ast$, as function of $x^\ast$ against the cumulative probability $P(x_i \leq x^\ast)$.
The conditional probability is equal to the cumulative in correspondence of the red dashed line.
Along these points, there is no attraction between schools of the same size.
This is not the case in both the two regions.}
	\end{figure}

In this last section, we bring more evidence on the effect of geography and comuni organization on the school-size by restricting our attention at two Italian regions: 
Abruzzo and Tuscany.
But same results stand by looking at regions with the same geographical features.
The two regions have very peculiar and representative geographical and administrative characteristics. 
Abruzzo is a mostly mountain region with a little flat seaside; 
it has four main head towns divided from each other by mountains. 
Conversely, Tuscany has many flat zones in the center and the mountain areas shape the region boundaries. 
Remarkably, it has a very high densely populated zone along the metropolitan area composed by Florence, Pisa and Livorno.

They also differ in terms of administrative organizations, Abruzzo favoring the establishment of comuni with a smaller size due to the presence of mountains.
As Fig. 6(a) makes clear, comuni distribute approximately as a log-normal pdf in both regions, i.e. as a parabola in a log-log scale (the blue line stands for Abruzzo pdf, the black for Tuscany).
Nevertheless, Tuscany has bigger cities.
The former region instead collects a larger number of small comuni that mostly do not provide schools.
We clusterize comuni using the algorithm in Eq. \ref{clustercomuni}.
The first bin collects comuni with a bit more than $100$ inhabitants.
They are $7$ in Abruzzo (none in Tuscany), none of them providing any school services.
The second bin gathers ten comuni in Abruzzo with $300$ inhabitants (none in Tuscany), of which only one has a school.
Comuni with about $600$ inhabitants are $40$ in Abruzzo and only $7$ in Tuscany.
	Only the $30\%$ of them has one school in Abruzzo, the $80\%$ in the latter region.
Overall, there are $53$ comuni in Abruzzo without schools;
only $3$ in Tuscany.

Such a differences reflects on the school-size distribution, depicted in Fig. 6(b).
Although primary schools distribute in both regions in terms of size with two peaks, both Abruzzo $m_1$ and $m_2$ are shifted on the left w.r.t. the Tuscany ones.
The average school-size is smaller in Abruzzo ($\hat{\mu}_{ABR} = 4.56$ ($\hat{\mu}_{ABR}/\ln(10) = 1.98$) versus $\hat{\mu}_{TOS} = 4.91$ ($\hat{\mu}_{TOS}/\ln(10) = 2.13$)), and, remarkably, the lower tail is fatter in the former region.
The cutoff for splitting the mixed distributions amounts to $128$ in Abruzzo and $151$ in Tuscany, and $31\%$ of the schools are clustered in the second peak in the former region;
$P(x_i > \bar{\mu}_{TOS}| \forall \, i \in TOS) = 0.38$ in the latter.

In Fig. 6(c) we show, following the same clustering technique used in Fig. 5(a), that the fraction of big schools within the comune $k$, $P_k(x_i > \bar{\mu}| \forall \, i \in k)$, increases with respect to the number of inhabitants in both regions, at least monotonically in comuni with a population smaller that $20$ thousands.
In this interval, a comparison with Italy figures, plotted in Fig. 5(a), reveals that both regions follow the same national pattern.
Yet, mountain regions, such as Abruzzo, have a significantly smaller concentration of big schools.
In particular, about $1/10$ comuni with just one school, gathered in the first bin on the left side, with an average population of roughly $2000$, have a school with more than $125$ students in Abruzzo.
In Tuscany, they are the $25\%$, about the same as national ratio.
In larger comuni, with an average population of $5000$ and two schools provided (the second bin), the probability of having big schools raises to $0.2$ in Abruzzo, still smaller than Tuscany where $\langle P_k(x_i > \bar{\mu}| \forall \, i \in k)\rangle_{h = 2} = 0.3$.

Small schools are mainly located in the countryside, and for that reason they are closer to each other in Abruzzo.
We investigate this point in Fig. 6(d), where we compute, and plot on the x-axis, the cumulative probability $P(x_i \leq x^\ast)$, for an arbitrary school size $x^\ast$, as function of $x^\ast$, and the correspondent conditional probability  $P(x_{\underline{t}} \leq x^\ast | x_i \leq x^\ast)$, on the y-axis, which is the fraction of smaller (than $x^\ast$) schools among the closest schools to a school of the same kind.
This quantity is equal to $74\%$ and $65\%$ for $x^\ast \equiv \bar{\mu}_{reg}$ in Abruzzo and Tuscany respectively, meaning that there is a greater probability that a small school matches with another of the same kind in the former region.
If the conditional probability were equal to the cumulative, as indicated by the red dashed line in Fig. 6(d), the sizes of neighboring schools would be independent.
This is not the case in either the two regions.
The probability that a small school has a smaller nearest neighbor is larger than the probability that any school is smaller than a given one.   
Indeed, the two curves (blue for Abruzzo and black for Tuscany) are significantly above the $45$ degree line for $P(x_i <x^\ast) < 0.6$ in Tuscany and for $P(x_i <x^\ast) < 0.7$ in Abruzzo. These probability values roughly correspond to the probabilities $P(x_i <\bar{\mu})$ in respectively Tuscany and Abruzzo, indicating that in both regions small schools are likely to belong to the small mountainous comuni, whose nearest neighbors are of the same class.   

\bigskip

\begin{figure}[!hbt]
\includegraphics[width=.99\textwidth]{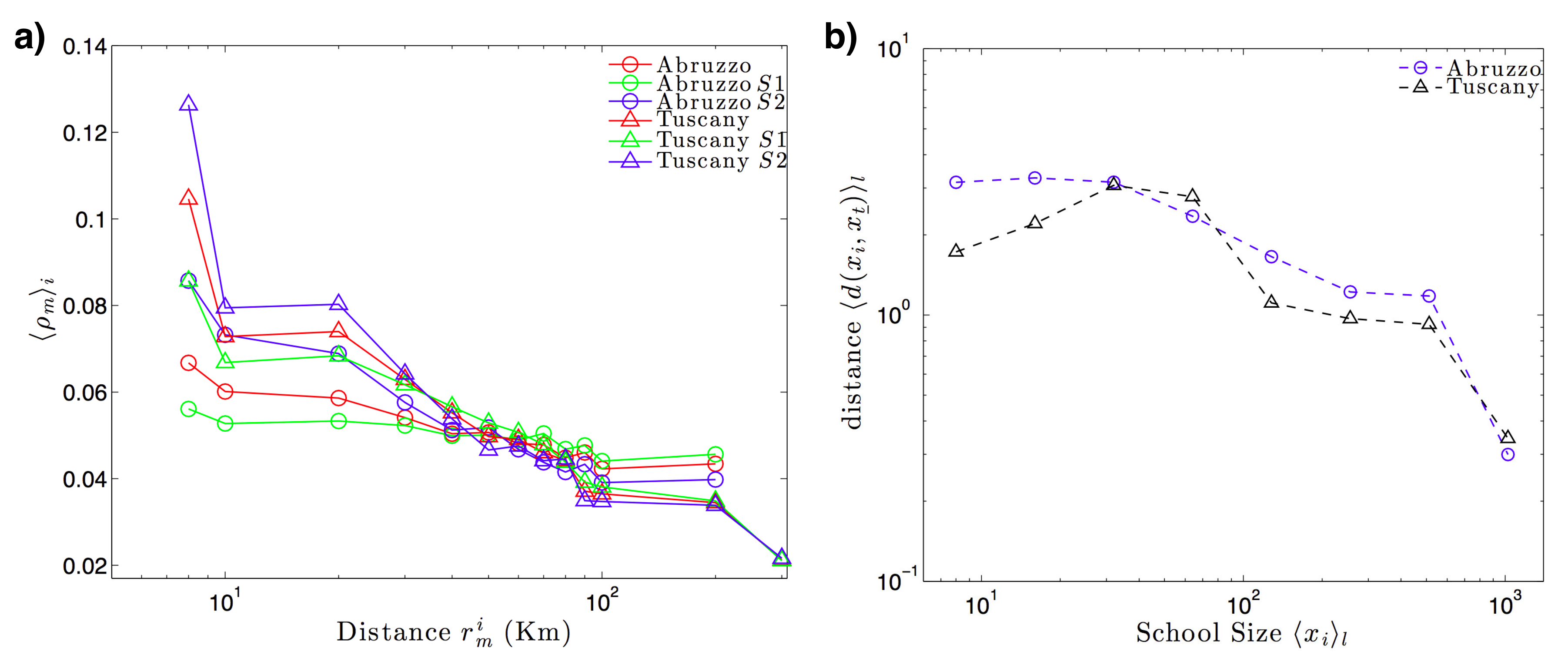}
\caption{  \textbf{Regional spatial analysis.} 
	\textbf{a.}
 $\langle \rho_{m}\rangle_i$ has been plotted, based on Eq. \ref{rho_i}, and \ref{rho_i_medio}, for the region of Abruzzo ($\square$) and Tuscany ($\triangle$).
	The red line draws the trajectory averaging among \emph{all} the schools in Italy.
Green and blue lines stand for small schools, i.e. $x_i \leq \bar{\mu}$, called $S_1$, and big schools, i.e. $x_i > \bar{\mu}$, called $S_2$, respectively.		
	\textbf{b.} 
The average distance, in km, between the closest schools, $\langle d(x_i, x_{\underline{t}})\rangle_l$, is plotted in Abruzzo (blue $\circ$) and Tuscany (black $\triangle$) with respect to the average size, $\langle x_i\rangle_l$.
Each cluster $l$ has been obtained by aggregating schools with near size according to Eq. \ref{clusterschool2}.
In Tuscany, the schools provided in small islands, at least $20 km$ far from the coast, have been removed in order to eliminate any artificial bias from the spatial analysis, whereas the $18\%$ of the schools, with no address provided in the MIUR dataset, have been geocoded in Tuscany according to the GPS localization of the city hall of the comune in which they stand.
The average distance between the closest schools decreases in both regions with respect to the average size meaning that, in general, small schools are more sparse than large schools that are more likely to be located in very dense zones, like cities.
}
	\end{figure}

We further study the attraction intensity among small schools by disentangling the effect between the countryside and dense zones.
To this end, we analyze the GPS location of the schools in the two regions and, for each school $i$, we compute the number of schools $n_m^i$ belonging within a circle of radius $r_m$ centered at each school $j$. 
We exclude from $n_m^i$ all the schools which do not belong to Tuscany or Abruzzo, respectively.  
To eliminate the effect of region's boundaries, we also compute areas $D_m^j$ as the areas of 
the intersections of these circles with a given region (Abruzzo or Tuscany).    
Thus $D_{m}^{i} \leq \pi (r_{m}^{i})^2$, because these areas do not include the seaside and administrative territories of other regions.
The difference between two subsequent circles yields the area of the annulus $A_{m}^{i} = D_{m}^{i} - D_{m-1}^{i}$.
The density of schools in the area $A_{m}^{i}$ is then defined as:
\begin{equation}
\label{rho_i}
\rho_{m}^{i} = \frac{n_{m}^{i} - n_{m-1}^{i}}{A_m^{i}},
\end{equation}
and the average density of schools as function of a distance to a randomly selected school is
\begin{equation}
\label{rho_i_medio}
\langle\rho_{m}\rangle_{i} = \frac{\sum_{N} n^{i}_{i} - \sum_{N} n_{m-1}^{i}}{\sum_{N} A_m^{i}}.
\end{equation}

In Fig. 7(a) red lines represent the average school-density around \emph{all} the schools in Tuscany and Abruzzo, which are 472 in the former and 1037 in the latter region.
Green lines describe the average school density around a small school
with $x_i \leq \bar{\mu}$, named $S_1$, whereas the blue lines describe
the density around large schools, $S_2$.  $64\%$ of the schools in
Abruzzo belong to the $S_1$ group, $53\%$ in Tuscany.
Fig. 7(a) collects evidence about the fact that small schools $S_1$ are located in low school density zones and, accordingly, have a smaller probability to be surrounded by competitor schools than large schools ($S_2$) located in densely populated areas.
In both regions, in fact, the green line goes under the blue one, for at least first $50 km$.  
In particular, within this distance, in Abruzzo the density stays almost constant at approximately $0.053$ meaning that $1$ school is provided every $20 km^2$.  
In Tuscany, this figure goes up to $0.07$, because of a generally higher population density, but yet small.

Fig. 7(b) confirms this pattern by showing that small schools have on average more distant nearest schools.
We look at the size of each school in both regions, and we define the geodetic euclidean distance between the school $i$ and the nearest $\underline{t}$ as $d(x_i, x_{\underline{t}})$.
A first look to the correlation coefficients reveals that the school size and this distance, $d(x_i, x_{\underline{t}})$, are negative correlated in both regions, but the magnitude is quite different, equal to $0.34$ in Abruzzo, that is $1.7$ times greater than in Tuscany ($0.20$).
To reduce the noise, we proceed by clusterizing schools according to their size.
The binning algorithm used is to base $2$:
\begin{equation}
	l = \{\forall \, i \in [1, \dots, N] : \, 2^{l-1} \leq x_{i} < 2^{l}\}.
	\label{clusterschool2}
\end{equation}
This clusterization yields $8$ bins, with different average sizes plotted on the x-axis of Fig. 7(b).
On the y-axis, we plot the average distance between the school $i$, that belongs to the bin $l$, and his nearest, i.e. $\langle d(x_i, x_{\underline{t}})\rangle_l$.
Each school-bin $l$ is depicted by blue circles for Abruzzo and black squares for Tuscany.
The average distance between the closest schools decreases in both regions with respect to the average size meaning that, in general, small schools are more sparse than large schools that are more likely to be located in very dense zones, like cities.
In Tuscany, the presence of schools within the hospitals plays an important role in keeping $\langle d(x_i, x_{\underline{t}})\rangle_l$ below $2 km$, for very small schools with less than $10$ students, whereas the schools provided in small islands, at least $20 km$ far from the coast, have been removed in order to eliminate any artificial bias from the spatial analysis.
The three first black bins are all below the blue ones, confirming, in accordance with the geographical features of the two regions, that  in Abruzzo small schools are more sparse and more likely to be located in the countryside where the school density is low (see Fig. 7(a)). 
Moreover, small schools on average have a distance to the nearest neighbor of $4-5 km$ which is the average distance between a small comune and a more school-dense one (see the Methods section).

\bigskip

The two regions then outline very different patterns of the school system in the countryside.  
In Abruzzo small schools are uniformly distributed across small comuni, as a result of a policy favoring the disaggregation of the comuni and school organization, due to a tight geographical constraint.  In Tuscany, instead, a different system has been implemented, according to geographic features and a higher population density, where small comuni are larger and do not necessarily have small schools, especially if they stand on very populated zones.

\section*{Discussion}

We have studied the main features of the size distribution of the Italian primary schools, including the sources of the bimodality, and we have investigated the relation with the Italian cities characteristics. 
The fat left tail of the distribution is the consequences of political decisions to provide small schools also in
small (mostly countryside) comuni, instead of increasing the efficiency of public transportations. 
This is most probably caused by the topographical features of the hilly terrain making transportation of students dangerous and costly. 
The evidence of this conclusions is that hilly cities like Palermo, Napoli, an, above all, Genoa, with steep mountains that end up into the see, have higher fraction of small schools than mainly flat cities like Torino and Milano.

The analysis of schools growth rates highlights that the schools dynamics follows the Gibrat law, and both the growth rate distribution and the size distribution are consistent with a Bose-Einstein process.
Alternatively, the exponential decay of the upper tail can be explained by a constraint by the size of the building or a traveling distance and transportation cost.

Despite our results are conducted using data on Italian primary schools, they predict that schooling organization would be different in another country with different geographical features.
Flat territory would lead to open schools in the main villages allowing the children residing in the smallest ones to travel daily.
This result is additionally supported by the fact that no territorial constraint has been imposed to the schooling choice.
Despite parents can enroll children in the most preferred school, primary students generally do not move \emph{across} comuni to attend a school.
Accordingly, we find that school density and school-size are prevalently driven by the population density and then by the geographical features of the territory, as a result of a random process in the school choice made by the parents.
This goes in the opposite direction with what has been found in other countries such as USA where school choices influence residential preferences of parents and drive the real estate prices in townships depending on the quality of their schools \cite{black99}.

The availability of new longitudinal school data will be relevant to a more in-depth analysis and further discussions. 
Moreover, the availability of data for other similar countries would favor comparison and would be useful to assert our theory.  
We believe that this study, and future research, can lead to a higher level of understanding of these phenomena and can be useful for a more effective policy making.

\section*{Methods}

\begin{figure}[!hbt]
\includegraphics[width=.99\textwidth]{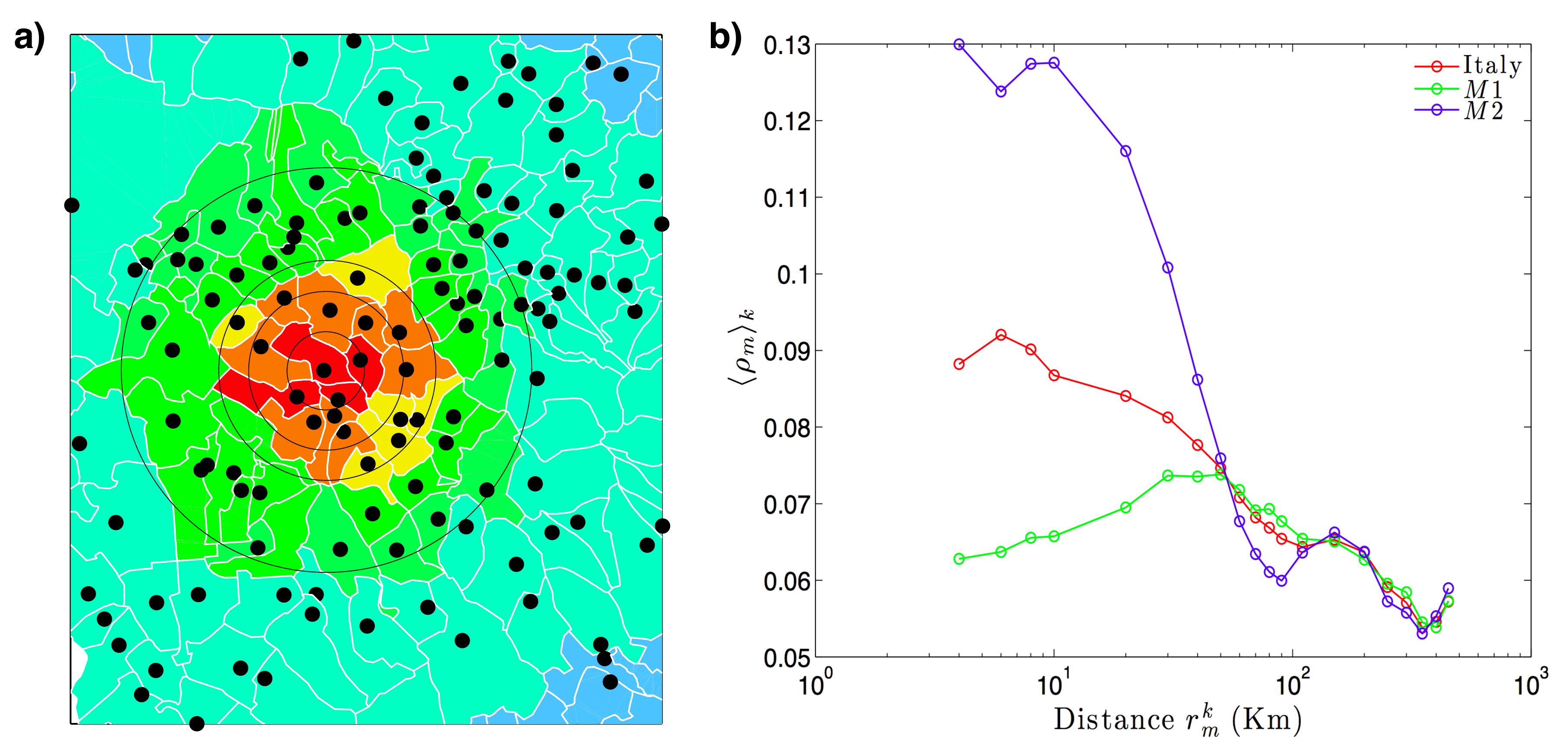}
\caption{  \textbf{Spatial analysis. a.}
Graphical example for a small comune in Abruzzo of the algorithm used in Fig. 8(b), based on the Eq. \ref{area_k}, \ref{rho_k}, and \ref{rho_k_medio}.
	Different comuni are colored according to the annulus in which they belong.
			\textbf{b.}
	$\langle \rho_{m}\rangle_k$ has been plotted for a radius $r^k_m$ of length $10^3$ across Italy.
	The red line draws the trajectory averaging among \emph{all} the cities in Italy.
	Green and blue lines stand for cities with probability $P_k (x_i > \bar{\mu}| \forall i \in k) \leq 1/2$, labeled $M1$, and $P_k (x_i > \bar{\mu}| \forall i \in k) > 1/2$, labeled $M2$, respectively. Maps generated with Matlab.
}
	\end{figure}
\bigskip

In this section we propose a novel algorithm for the analysis of spatial 
distribution of primary schools in entire Italy. This algorithm is needed
if the exact coordinates of individual schools are not available, but instead, the centers and the territories of all the communi are known. 
For each commune $k$, we define a gravity center $g_k$ of its territory corresponding to the GPS location of its city hall, and $t_{k}$ as the area of the comune administration.  
In Italy the city hall is located in the center of the densely populated part of the administrative division, in order to be easily reachable by the majority of inhabitants.  
We develop a novel spatial-geographical approach consisting of a sequence of geographic regions bounded by two concentric circles, that we exemplified in Fig. 8(a) for a comune in Abruzzo.  
First we define a set $Z^k_{m}$ of comuni whose city halls are within a circle of radius $r_{m}^{k}$ and the center at the city hall of comune $k$. 
Formally, 
\begin{equation}
Z^k_{m}=\{\forall\ j \in [1,..,K]: d(g_{k},g_{i}) \leq r_{m}^{k}\}.
\end{equation}
Next we compute the number of schools provided by the comuni which are members of set $Z^k_{m}$ that is defined by 
\begin{equation}
n_{m}^{k}=\sum_{j\in Z_{m}^{k}}n_j
\end{equation}
and their area
\begin{equation}
\label{area_k}
 D_{m}^{k}=\sum_{j\in Z_{m}^{k}} t_{j},
\end{equation}
where $t_j$ is the area of comuni $j$.
Next we compute the area associated with all the comuni in the $m$-th concentric annulus surrounding comune $k$ as the difference between the area associated with the larger circle $m$ of radius $r_{m}^{k}$ and the area associated with the smaller circle $m-1$ of radius $r_{m-1}^{k}$, i.e. $A_{m}^{k} = D_{m}^{k} - D_{m-1}^{k}$.
In Fig. 8(a), each comune territory is colored with different colors according to the annulus in which they belong.

The density of schools in the area $A_{m}^{k}$ is then defined as:
\begin{equation}
\label{rho_k}
\rho_{m}^{k} = \frac{n_{m}^{k} - n_{m-1}^{k}}{A_m^{k}}
\end{equation}
Then we compute the average density of schools around any school in Italy as:
\begin{equation}
\label{rho_k_medio}
\langle\rho_{m}\rangle_{k} = \frac{\sum_{K} n^{k}_{m} - \sum_{K} n_{m-1}^{k}}{\sum_{K}A_m^{k}}
\end{equation}

In Fig. 8(b), we plot $\langle\rho_{m}\rangle_{k}$ averaged over all the $K = 8092$ Italian comuni as a function of the radius $r_m$ that goes up to $10^3$ Km across the entire Italy.  
The red line represents the average school-density among \emph{all} the cities in Italy.  
On average, Italian comuni stand within very dense zones providing almost $1$ school per $10 km^2$.  
The dense zones generally last for $10 km$ and, after that, a smoothed depletion zone is experienced.
However, the average distance between a comune $k$ and a very large city with many schools is about $100km$, accordingly we see a second peak in the average school density at distance $100km$.

The full sample analysis basically averages heterogeneous characteristics that feature different types of comuni.  
The interaction among schools can be better understood by splitting the sample according to $P_k (x_i > \bar{\mu}| \forall i \in k)$.  
In Fig. 8(b), comuni with $P_k (x_i > \bar{\mu}| \forall i \in k) \leq 1/2$, i.e. with predominantly small schools, are named $M1$.  
The others, with predominantly big schools, are called $M2$.
\begin{itemize}
	\item $M2$-comuni, the blue line, are (on average) more likely to be surrounded by school-dense cities.
They are cities located in densely populated areas (depicted in red in Fig. 4(d)) where the school density is large (1.3 schools stand on average within $10 km^2$).
As far as the distance increases mountainous areas (and hence $M1$-comuni) are encountered and, as a result, the density of schools is found to dramatically decrease.   
	\item The green line describes instead cities labeled $M1$ where a smaller school density is found.
Within $10 km$, in fact, almost $1$ school every $20 km^2$ are encountered on average, about the half of what we find for the $M2$-comuni. 
This is because $M1$-comuni mainly stand along the countryside (those depicted in blue in Fig. 4(d)) where school density slowly increases with distance and reach a maximum at approximately $40km$, which can be interpreted as a typical distance to a densely populated area in a neighboring mountain valley. After this
distance the density of schools around $M1$ and $M2$ comuni behave approximately in the same way. 
\end{itemize}

\section*{Supplementary}

\subsection*{Italian private primary schools versus public primary schools: a comparison.}

\bigskip

In the paper we addressed the source of the bimodality by considering all the Italian primary schools. 
Here we focus on the potential effect of school type on the school-size distribution. 
Our dataset collects $N = 17,187$ primary schools in Italy. 
The fraction of private schools was always low during the past century. 
In Italy only the $9\%$ of the total of primary school are private. 

The main source of primary school privatization within the country is religion.
Most of the private schools are venues where education is strictly connected with the Catholic confession. 
Among the private schools more than $73\%$ are of Catholic inspiration. 
Straightforward historical roots are expected to explain the location of the Italian Catholic private schools and only marginal are the geographical reasons: 
private schools are in fact only the $6.54\%$ of the mountain schools.

We define $\mathcal{M}$ the set of comuni $k$ that are in mountains that, according to the Law n. 991/1952, are those that have at least the $80\%$ of their territories above the $600$ meters above the sea and an altitude gap between the higher and the lower point not least than $600$ meters.
Each comune $k$ has $n_k$ schools and a fraction of private schools in this comune defined as $P(i \in \mathcal{P} | \forall i \in k) \equiv \eta_k$, where $i$ is the school ID. We also define the school-size of a private school $i$ that resides in a mountain comune as $x_{i \in \mathcal{P}, \mathcal{M}}$.
Analogously, $x_{i \in \mathcal{\bar{P}}, \mathcal{\bar{M}}}$ stands for the size of a public school residing in a non-mountain comune.

\begin{figure}[!ht]
	\centering
		\includegraphics[width=.99\textwidth]{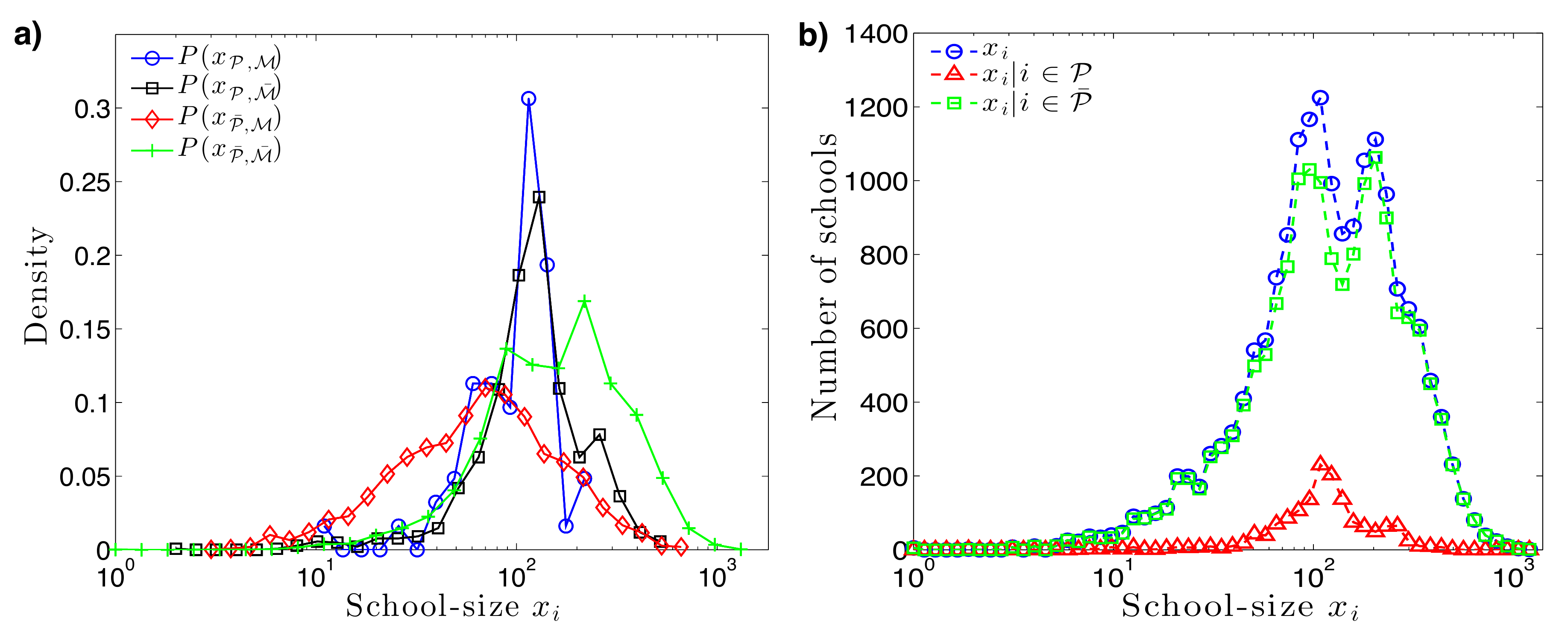}
		\caption{\small{\textbf{a.} Italian primary school-size distribution disentangled by school type (private, $\mathcal{P}$, versus public, $\mathcal{\bar{P}}$) and geography (mountain, $\mathcal{M}$, versus non-mountain, $\mathcal{\bar{M}}$).
\textbf{b.} Italian primary school-size distribution by school-type.
The blue pattern replicates Fig. 1a in the main text. 
}}
\end{figure}

Figure 9(a) shows that neither private mountain schools $(\mathcal{P}, \mathcal{M})$ nor private schools that reside in flat territories $(\mathcal{P}, \mathcal{\bar{M}})$ seem to contribute significantly to the left tail of the school-size distribution. 
Both the ($\circ$) blue and the ($\square$) black lines, respectively, depict two relatively narrow school-size distributions around $100$ students per school, the ($+$) green $(\mathcal{\bar{P}}, \mathcal{\bar{M}})$ and the ($\diamond$) red lines $(\mathcal{\bar{P}}, \mathcal{M})$. 
In accordance with the results shown in the main text, mountain public schools mostly contribute to the left tail of the distribution. 
Finally, the distributions of private schools both for mountain and flat regions are almost identical even though there are only 449 mountain private schools and one might expect large statistical uncertainty. 

Figure 9(b) draws the school-size distribution without considering geography but only distinguishing with respect to the school-type.
Frequencies are then shown for private (red $\triangle$) and public (green $\square$) schools and compared with the distribution of all the Italian primary schools (in blu $\circ$) that replicates Figure 1a in the main text.
It confirms that private schools play only a slight role in generating the left peak that yet remains even conditioning by school type.

Figure 10(a) plots the fraction of private schools in each bin $c$ of comuni with given altitude, $\eta_k$, against their altitude above the sea level, $\chi_{k}$. 
In order to reduce the noise, we binned comuni according to the altimetry:
\begin{equation}
	c = \{\forall \, k \in [1, \dots, K] : \, 2^{c-1} < \chi_{k} \leq 2^{c}\}.
	\label{clustercomuni}
\end{equation} 
It yields 11 bins, $c \in [1, \dots, 11]$, each of them collecting comuni according to the meters above the sea level.
Overall, the figure provides evidence of negative correlation between the fraction of private schools and the altitude above the sea of that comune (in $\ast$ blue), in contrast with the fraction of schools (both private and public) in the bin which follows an hill shaped relationship (in $\square$ magenta).
Therefore, we conclude that there are relatively more private schools in the flat zones with respect to mountains.

\begin{figure}[!ht]
	\centering
		\includegraphics[width=.99\textwidth]{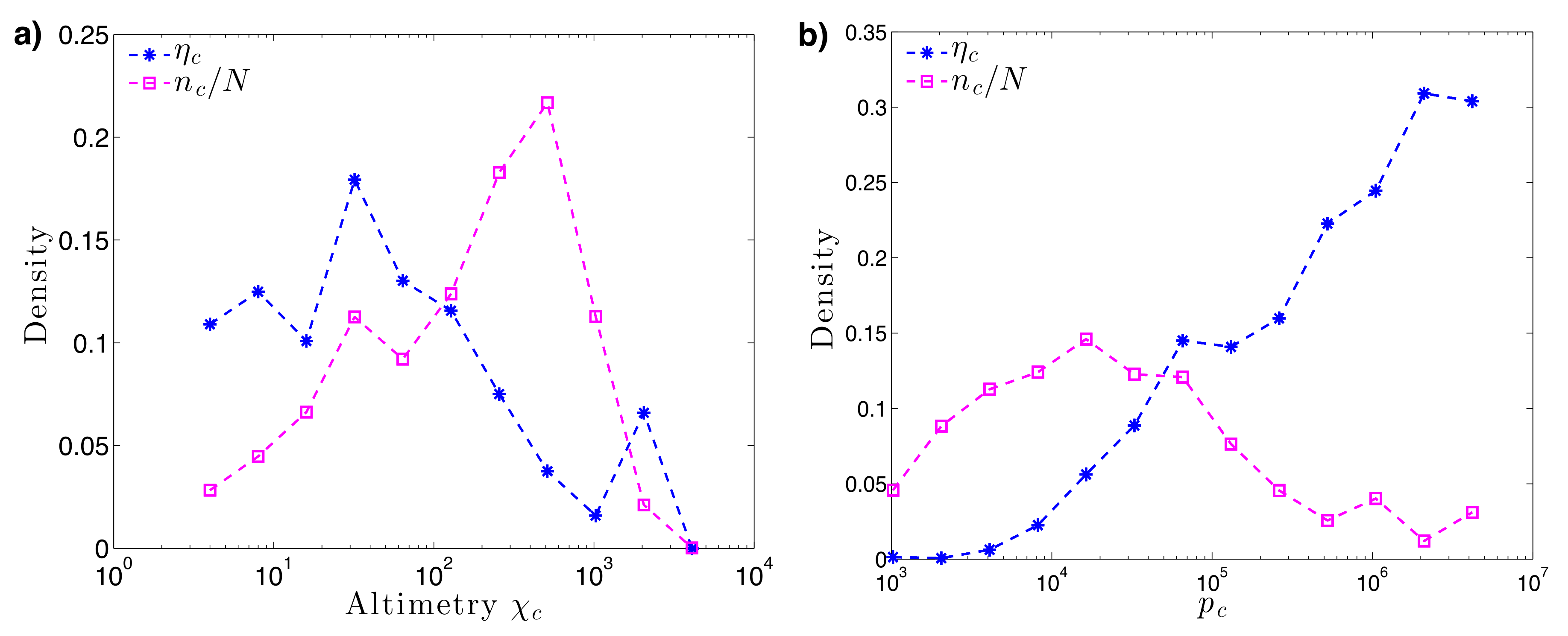}
		\caption{\small{
\textbf{a}. Correlation between the fraction of private schools in the bin $c$, $P(i \in \mathcal{P} | \forall i \in c) = \eta_c$ (in $\ast$ blue), and the altitude above the sea of that bin, $\chi_{c}$.
The same relation is depicted in ($\square$) magenta for the fraction of schools (both private and public), $n_c/N$.
\textbf{b}. Correlation between the fraction of private schools in the bin $c$, $P(i \in \mathcal{P} | \forall i \in c) = \eta_c$, drawn in ($\ast$) blue, and the number of inhabitants in that bin, $p_c$.
As a robust check we also plot in ($\square$) magenta the fraction of schools (both private and public) in each bin c, $n_c/N$, versus the population.
}}
\end{figure}

Finally, using the same binning algorithm in Eq. 4, Figure 10(b) shows strong positive correlation between the fraction of private schools in the bin $c$, $\eta_c$, and the number of inhabitants in that bin, $p_c$ (in $\ast$ blue), confirming that the location of the Italian Catholic private schools mainly roots in the more populated comuni.
As a robust check we also plot in ($\square$) magenta the fraction of schools (both private and public) in each bin $c$, $n_c/N$, that, consistently with the analysis run in Fig 3 in the main paper, approximates the Italian population distribution with a slight skewed shape. 
The two lines differ remarkably.
In very small comuni ($p_c < 10^4$), where a greater quantity of schools is provided, we count a tiny fraction of private ones.
Conversely, in the biggest comuni data show that the relation goes the other way around with a definite bigger fraction of private schools provided (e.g. in Rome $\eta_k \approx 0.30$).

Big flat comuni are then very likely to be the places where mostly private Italian primary schools are located around the country.
We conclude that privatization has been driven across the years for religious confessional purposes rather than following the unmatched education demand in the countryside due to the lack of the public system.

\subsection*{Testing unimodality in the school-size distributions of flat comuni.}
\bigskip

In this section we address concerns on bimodality on the school-size distribution of flat comuni.
In the main text we have demonstrated that geography is the main source of bimodality in the school-size pdf showing that mountain schools clusterize around $m_1$.
Yet there might be other confounding factors that might keep a second peak, i.e. $m_1$, in the school-size pdf of the schools that reside in flat comuni.

In Fig. 4b we distribute schools according to the number of students, $x_i$, conditional on the altimetry of comuni.
As we discuss in the main text (see Section \emph{City-size and the school-size}) this exercise gives five distributions that translate according to the height (location effect).
The pdfs of mountain schools stand on the left and on the right we have flat schools.
The ($\circ$) green line shows the school-size distribution for $N_{250m} = 3,033$ schools that reside  in comuni with around 250 meters from the sea level.
Despite the pdf does not show a clear kinky distribution that converges to $m_2$, that potentially might show bimodality, here we demonstrate that \emph{statistically} the hypothesis in favor of unimodality can not be rejected.

To see that we use the complementary error function to estimate the probability that the number of schools in the central bin $n_1$ is not significantly smaller than and the number of schools in the next two bins $n_2,n_3$ are not significantly larger than a certain number $n^\ast$ provided that the standard deviation of the number of schools in these bins due to small statistics is $\sqrt{n^\ast}$:
\begin{equation}
	p(n^*) = \frac{1}{2} \Pi_i \text{erfc}\bigg(\frac{|n_i - n^*|}{\sqrt{2n^*}}\bigg)
\end{equation}
This is equivalent to test the hypothesis that the distribution is unimodal. 
In the school-size distribution for schools that reside in comuni with around 250 meters above the sea, the central bin collects $n_1 = 639$ schools.
On either sides there are two other bins that collect $n_2 = 670$ and $n_3 = 646$ respectively.
The probability that the distribution is not bimodal is maximum for $n^* = 646$ where it is equal to $p_{max}(n^* = 646) = 0.15$.
Fixing a level of confidence of $0.10$ therefore we cannot reject the hypothesis of unimodality.

\section*{Acknowledgements}
We acknowledge Stefano Sebastio for many interesting and important discussions.
We are also grateful to all the participants at the LIME seminars in IMTLucca.

\section*{Author contributions}
A.B. \& R.D.C. analyzed the data. R.D.C. created the maps. A.B., R.C.D. \& S.B. devised the research, wrote and revised the main manuscript text.

\section*{Additional Information}
Competing financial interests: The authors declare no competing financial interests.


\begin{thebibliography}{}

\bibitem{gabaix09}
	Gabaix, X., 
	Power Laws in Economics and Finance.,
	\emph{Annu. Rev. Econ.}
	\textbf{1},
	255--93,
	(2009).
	
\bibitem{gabaix99}
	Gabaix, X.,
	Zipf's Law for Cities: An Explanation.,
	\emph{Q J Econ.}
	\textbf{114},
	739--67,
	(1999).  
	
\bibitem{allen1997cities}
	Allen, P.M.,
	\emph{Cities and regions as self-organizing systems: models of complexity},
	(Routledge, 1997).

\bibitem{amaral98}
	Amaral, L. A. N., et al.,
	Power Law Scaling for a System of Interacting Units with Complex Internal Structure.,
	\emph{Phys. Rev. Lett..}
	 \textbf{80},
	 1385--1388
	 (1998).

\bibitem{byrne2002complexity}
	Byrne, D.,
  	\emph{Complexity theory and the social sciences: an introduction},
	(Routledge, 2002).
	
\bibitem{caves98}
	Caves, R. E.,
	Industrial Organization and New Findings on the Turnover and Mobility of Firms.,
	\emph{J Econ Lit.}
	\textbf{36},
	1947--82,
	(1998).
	
\bibitem{bak1997nature}
	Bak, P.,
	\emph{How nature works},
	(Oxford University Press, 1997).
	
\bibitem{kauffman1996home}
	Kauffman, S.,
 	 \emph{At Home in the Universe: The Search for the Laws of Self-Organization and Complexity: The Search for the Laws of Self-Organization and Complexity.},
  	(Oxford University Press, 1996).

\bibitem{decret1}
	Gazzetta Ufficiale,
	\emph{Decreto del Presidente della Repubblica del 20 marzo 2009 n. 81},
	(2 luglio 2009).

\bibitem{decret2}
	Disposizioni concernenti la riorganizzazione della rete scolastica, la formazione delle classi e la determinazione degli organici del personale della scuola.,
	\emph{Decreto Ministeriale}
	\textbf{331},
	(24 luglio 1998).

\bibitem{Belmonte2013}
	Belmonte, A. \& Pennisi, A., 
	Education reforms and teachers needs: a longterm territorial analysis.,
	\emph{IJRS.} 
	\textbf{12}, 
	87-114
	(2013).

\bibitem{dewit05}
	De Wit, G.,
	Firm Size Distributions: An Overview of Steady-State Distributions Resulting form Firm Dynamics Models.,
	\emph{Int J Ind Organ.}
	\textbf{23},
	423--50
	(2005).

\bibitem{fu2005growth}
	Fu, D. et al.,
	The growth of business firms: Theoretical framework and empirical evidence.,
 	\emph{Proc Natl Acad Sci USA.}
	\textbf{102},
 	18801--18806
	(2005).

\bibitem{decret3}
	Gazzetta Ufficiale,
	\emph{Decreto del Presidente della Repubblica dell'8 marzo 1999 n. 275}, 
	(10 Agosto 1999).
	
	
\bibitem{pitman1939estimation}
	Pitman, E. J. G.,
  	The estimation of the location and scale parameters of a continuous population of any given form.,
  	\emph{Biometrika}
	\textbf{30},
 	391--421
	(1939).
	
	
\bibitem{stanley95}
	Stanley, M. H. R. et al.,
	Zipf plots and the size distribution of firms.,
	\emph{Econ Lett.}
	\textbf{49}, 
	453--457
	(1995).
	
	
\bibitem{gibrat}
	Gibrat R.,
	\emph{Les In\'egalit\'es \'economiques},
	(Recueil Sirey 1931).
	
\bibitem{sutton97}
	Sutton, J.,
	Gibrat's Legacy.,
	\emph{J Econ Lit.}
	\textbf{35},
	40--59
	(1997).
		
\bibitem{fu2007generalized}
  	Fu, D. et al.,
	A Generalized Preferential Attachment Model for Business Firms Growth Rates-I. Empirical Evidence.,
	\emph{Eur Phys J B.}
	 \textbf{57},
	127--130
	(2007).

\bibitem{pammolli2007generalized}
	Pammolli, F. et al.,
	A generalized preferential attachment model for business firms growth rates: II. Mathematical treatment.,
	\emph{Eur Phys J B.}
	\textbf{57},
	131--138
	(2007).
	
\bibitem{axtell01}
	Axtell, R. L., 
	Zipf Distribution of U.S. Firm Sizes.,
	\emph{Science}
	\textbf{293},
	1818--20
	(2001).

\bibitem{growiec07}
	Growiec, J., Pammolli, F., Riccaboni, M. \& Stanley, H.E., 
	On the Size Distribution of Business Firms.,
	\emph{Econ Lett.}
	\textbf{98},
	207--12
	(2007).
	
\bibitem{stanley96}
	Stanley, M. H. R., et al.,
	Scaling behaviour in the growth of companies.,
	\emph{Nature}
	\textbf{379},
	804--6
	(1996).
		
\bibitem{ayebo2003asymmetric}
	Ayebo, A. \& Kozubowski, T.J.,
	An asymmetric generalization of Gaussian and Laplace laws.,
  	\emph{J. Probab. Stat. Sci.}
	\textbf{1},
	187--210
	(2003).


\bibitem{bookbul}
	Buldyrev, S. V., Pammolli, F., Riccaboni, M., \& Stanley, H. E.,
	\emph{The Rise and Fall of Business Firms},
	To be Published (2014).		

\bibitem{wang09}
	Wang, J., Wen, S., Symmans, W. F., Pusztai, L. \& Coombes, K. R., 
	The bimodality index: a criterion for discovering and ranking bimodal signatures from cancer gene expression profiling data., 
	\emph{Cancer Inform} 
	\textbf{7},
	199--216
	(2009).
	
\bibitem{lauset2009power}
	Clauset, A., Shalizi, C. R. \& Newman, M. E. J.,
  	Power-law distributions in empirical data., 
 	 \emph{SIAM review.}
	 \textbf{51},
 	661--703
	(2009).
	
\bibitem{eeck04}
	Eeckhout, J.,
	Gibrat's Law for (All) Cities.,
	\emph{Am Econ Rev.}
	\textbf{94},
	1429--51
	(2004).

\bibitem{reed01}
	Reed, W. J., 
	The Pareto, Zipf and Other Power Laws.,
	\emph{Econ Lett.}
	\textbf{74},
	15--9
	(2001).
	
\bibitem{black99}
	Black, S. E.,
	Do better schools matter? Parental valuation of elementary education.,
	\emph{Q J Econ.}
	\textbf{114},
	577-599
	(1999).

\end{thebibliography}
\end{document}